# Ice-templating beet-root pectin foams: controlling texture, mechanics and capillary properties

*Sarah Christoph[a], Ahmed Hamraoui[a,b], Estelle Bonnin[c], Catherine Garnier[c], Thibaud Coradin[a], Francisco M. Fernandes[a]\**

[a] Sorbonne Université, CNRS, Laboratoire de Chimie Matière Condensée de Paris, LCMCP, F-75005 Paris, France

[b] Université Paris Descartes, Faculté des Sciences Fondamentales et Biomédicales, 45 rue des Saints-Pères, 75006 Paris, France

[c] INRA UR 1268 - Biopolymères - Interactions - Assemblages, BP 71627, 44316 Nantes Cedex 03, France

**Abstract**

Sugar beet pectin is a byproduct of the sugar industry with a particularly low gelling power which hinders its application as gelling agent and thickener. Here we consider the use of freeze casting to shape sugar beet pectin into lightweight foams. Freeze casting processing conditions such as the applied thermal gradient and the polysaccharide concentration were explored to obtain macroporous lightweight foams with different textures. The precise control over the foams' texture and pore anisotropy was decisive for their performance as liquid transport devices by capillary ascension and for their mechanical performance. Overall, the obtained results show that the formation of highly anisotropic structures using freeze casting can be instrumental in the upcycling of polysaccharide industrial byproducts.

*Corresponding author: francisco.fernandes@upmc.fr

# 1. Introduction

Macroporous materials are omnipresent in everyday life. From natural examples such as trabecular bone, balsa wood or cork [1] to man-made materials used in transport applications [2], antifrost surfaces [3], thermal insulation materials [4] or water harvesting devices [5], we are surrounded by outstanding, lightweight, functional materials. To balance functionality and apparent density, Nature has relied on highly specific – often cell-mediated – bottom-up processes to precisely shape the materials' porosity. Among these materials, a small portion, such as plant stems and tree trunks [6] or porcupine quills [7] displays highly anisotropic macropores. Two main reasons lie behind these aligned structures: mechanical performance [8–11] and/or fluid transport [12,13]. Several processing strategies are available to replicate such highly controlled porous structures [14–17], however few are as straightforward as freeze-casting, a technique based on oriented ice templating that has gained considerable during the last decade [18]. Freeze-casting relies on the use of ice crystal growth to create porosity. The method presents the advantage of being easy to implement, cost effective and applicable to a wide range of materials [19]. In particular, since the technique relies on the use of low temperatures, it can be particularly relevant in the processing of materials prone to thermal denaturation such as biopolymers. The technique was first described in 1954 [20] for the processing of refractory powders and has been used since to shape ceramics [21], metals [22,23], clays [24], polymers [25,26] and composite materials [27,28].

Pectin is a polysaccharide present in plant cell wall that finds extensive application as gelling and thickening agent [29]. The gelling properties of this heteropolysaccharide are strongly dependent on its source and extraction conditions [30]. Pectin from sugar beet – a source yielding pectin with particularly low gelling power – displays fewer direct industrial application without extensive chemical modification than high gelling ability sources such as citrus peel or apple pomace [31]. Such difference in gelling power is mainly due to the presence of acetyl esters on the galacturonic acid chain [32]. As a result, sugar beet pectin remains a largely undervalued material, often considered as a byproduct of the sugar industry with few valorization alternatives beyond the production of arabinose and galacturonic acid [33]. Designing new processing approaches that enable the application of sugar beet pectin represents an opportunity to upcycle a largely overlooked industrial resource. Here we report how ice templating techniques can shape pectin to obtain biobased materials with adjustable porosity. We demonstrate that control over pore size and pore morphology can provide pectin-based materials with interesting mechanical and liquid transport properties. In summary, the finely controlled porous structure imposed during the ice templating unveils the possibility to apply a wide variety of water-soluble industrial sub-products for their potential application as lightweight structural elements, insulation or liquid transport devices.

## 2. Material and Methods

### 2.1. Materials

Sugar beet pectin powder was obtained from CP Kelko and used without further purification. Degrees of methylation and acetylation were 49.00 and 20.90 respectively. The detailed composition is presented in Supplementary Information (Table S1). Pectin aqueous solutions ranging from 10 to 50 g.L$^{-1}$ were prepared by dissolving the pectin powder in ultra-pure water followed by stirring under magnetic agitation for 24 hours at room temperature.

### 2.2. Process

The macroporous pectin monoliths were shaped by growing ice crystals into aqueous pectin solutions, followed by freeze-drying in a Christ Alpha 2-4 LD freeze-dryer operating at 0.06 mbar for a 48 hour period, in order to sublimate the ice crystals.

Four freezing conditions were investigated: two using laboratory freezers (-20°C and -80°C), one using a liquid nitrogen (LN) bath and one using a custom-made freeze-casting setup (FC) (Figure S1). The samples frozen in the laboratory freezers (4 mL at 40 g.L$^{-1}$) were placed in PE cylindrical molds ($\varnothing$ = 19 mm) and left overnight in the freezer either at -20°C or -80°C before freeze-drying. Samples obtained through ice templating in LN were first poured in similar 19 mm diameter molds at room temperature and subsequently plunged for 5 minutes into LN before freeze-drying. Various samples were obtained by freeze-casting, using a previously reported setup [34]. The device consists in a heat conductive copper rod partially immerged into LN. Temperature profiles at the top of the rod (interface between sample and copper rod) is set through a heating resistance controlled by a dedicated PID. The pectin solutions (3 mL) were placed at the top of the copper rod in 15 mm diameter cylindrical molds. The samples were then cooled down from 20 °C to -60°C at a controlled cooling rate of 10°C.min$^{-1}$. To ascertain the impact of the pectin solution concentration on the final material properties, samples were prepared from initial pectin solutions ranging from 20 g.L$^{-1}$ to 50 g.L$^{-1}$.

### 2.3. Characterization

The cooling profiles for each freezing process were determined by inserting a K thermocouple at the center of 40 g.L$^{-1}$ pectin solutions before freezing. The temperature profiles for the samples frozen inside the freezers (-20°C and -80°C) and by plunging into LN were recorded until thermalization. For the freeze-cast samples, the temperature profile was determined at the center of the pectin solution and at the interface between the copper rod and the pectin solution.

Samples obtained after freeze drying were manually cut with new scalpel blades. Scanning electron microscopy observations were performed on Hitachi S-3400N SEM. The samples were sputter coated with 20 nm of gold and observed under 3-4 kV acceleration and 30 µA probe current. SEM pictures were analyzed using the OrientationJ plugin [35] in ImageJ software for pore morphology characterization. The foams' pore dimensions and pore wall thickness were determined on 150 and 25 individual measurements, respectively.

Wetting behavior of the foams was assessed by capillary rise of a solution of Disperse Red at 0.2 gL$^{-1}$ in ethanol. Cylindrical pectin foams (triplicates with height between 6 and 8 mm) were placed above a Disperse Red solution and the impregnation recorded on camera at 30 fps. The resulting footage was analyzed using the "Reslice" operation to deduce ascension profiles in ImageJ software. The surface tension of Disperse Red in ethanol was measured using the Wilhelmy plate method on a Krüss K11model.

Mechanical behavior under compression was assessed using an Instron 5965 universal testing machine equipped with a 50 N load cell. Samples were cut into 1 cm$^3$ cubes and compressed between stainless steel plates up to 50% strain at constant displacement rate of 1 mm.min$^{-1}$. The corresponding stress-train curves were obtained for 5 replicas per sample.

**3. Results and discussion**

3.1. From freezing to freeze-casting pectin foams

Self-supporting, low density macroporous biopolymer materials were obtained by ice-templating pectin solutions. Ice templating is particularly adapted to the processing of biopolymers that are prone to thermal degradation such as pectin since the processing steps are carried out below ambient temperature. In addition, the technique does not require leaching with possible denaturating solvents to unveil the porosity as required for other porogens such as latex beads. Since most chemical species are largely insoluble in ice, an increase in concentration of the solutes occurs in the interstices formed between the growing ice crystals. When ice crystallization stops due to the lack of available water to freeze, the biopolymer located in between the ice crystals is structured into a solid macroporous network. The biopolymer foam obtained after ice sublimation therefore corresponds to the negative imprint of the formed ice crystals.

Different freezing conditions were explored to highlight the relationship between the conditions of ice formation and the final morphology of the porous monolith. Figure 1 displays the temperature profile at the core of the samples during freezing. The first three conditions (freezers at -20°C, -80°C and LN bath) differ in the target temperature (-20°C, -80°C or -196°C respectively) but the orientation of the temperature gradient within the samples is similar. In the case of freeze-casting however, multiple parameters differ. This setup allows controlling the freezing conditions both spatially (a specific temperature gradient is applied thanks to the marked difference in thermal conductivity between the copper rod and the surrounding sample environment) and in time (cooling rate can be tuned between 1 and 10 °C.min$^{-1}$).

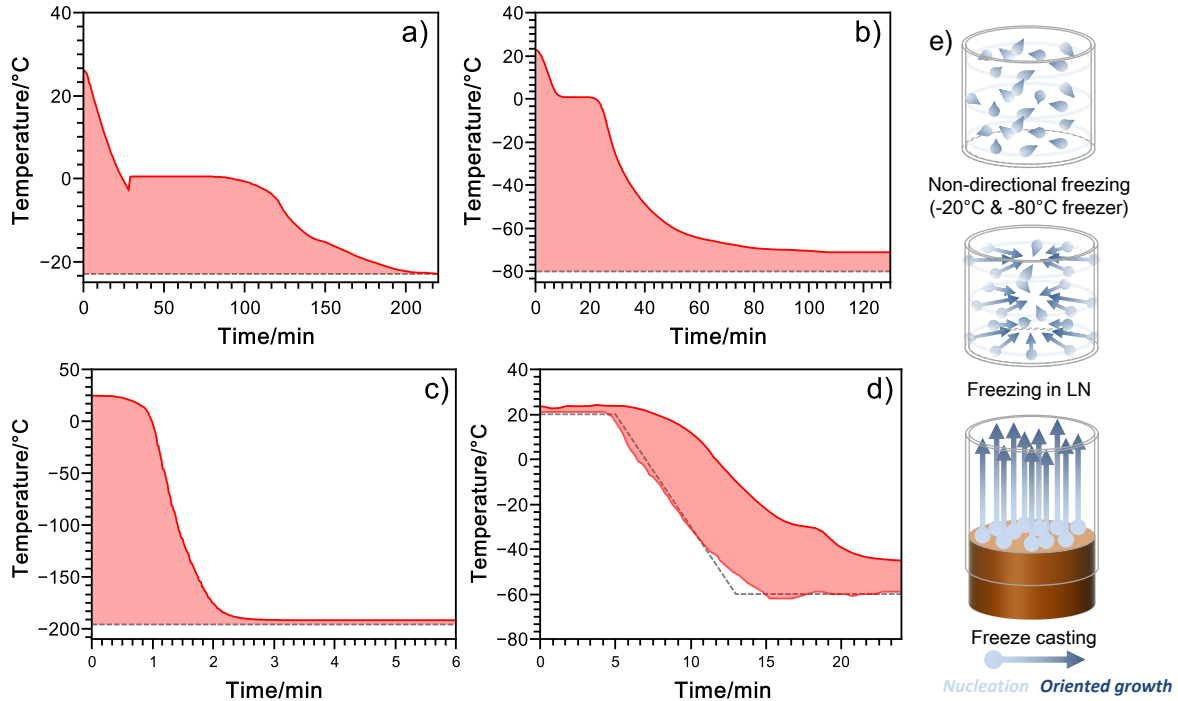

**Figure 1.** Temperature profiles during freezing of 40 g.L$^{-1}$ pectin solutions using different ice templating conditions, a) -20°C freezer, b) -80°C freezer, c) liquid nitrogen (LN,-196°C), d) freeze casting at 10°C.min$^{-1}$ (-60°C). Full red line is the temperature at the core of the sample, full pale pink line is the temperature at the interface between the copper rod and the sample (for the freeze casting setup) and dashed line represents the target temperature. d) Schematic representation of ice growth for the different freezing processes.

Samples frozen in the laboratory freezers display comparable cooling temperature profiles. After an initial cooling period, the temperature remains constant during the growth of the ice crystals in the whole sample. The temperature then decreases to reach the target temperature. At lower target temperature (-80°C) the initial cooling rate is higher (3.6 °C.min$^{-1}$ in the -80°C freezer *vs* 1.2 °C.min$^{-1}$ in the -20°C freezer), and the ice growth time reduced (around 10 min at -80°C *vs* 50 min at -20°C). The sample frozen at -20°C shows a small supercooling effect reaching -2.7°C before rising back to 0°C during the freezing step. By dipping in LN, no isothermal phase transition can be observed, but a significant delay in the initial cooling occurs.

Using the freeze casting setup, the difference between the target temperature and lower boundary temperature is minimal throughout the whole freezing process. At the center of the sample (Figure 1d) the temperature delay with respect to the cooling program reflects the unidirectional temperature gradient created with the freeze casting setup. The differences in the temperature profiles, and how these translate into different nucleation and growth kinetics of the ice crystals, can be linked to the final morphology of the porous materials.

Figure 2 presents the micro- and macroscopic morphology of foams obtained in the different freezing conditions. For each sample a cross section and a longitudinal section at different length scales are presented. Samples obtained at -20°C and -80°C present similar morphologies in both orientations. In contrast, the sample prepared in LN shows a strong pore anisotropy. These observations are confirmed by orientation mapping (Figure S2). The pores' morphologies can be related to the ice crystals nucleation and growth. As indicated in Figure 1, the presence of moderate initial cooling rates and a temperature plateau for the samples prepared at -20°C and -80°C show that there is no or limited temperature gradient inside the sample. The cooling rate is slow enough for the sample temperature to remain homogeneous throughout the whole sample. As a result, the whole sample reaches transition temperature at the same time, resulting in simultaneous nucleation of ice in the whole volume. During the following ice growth phase there is limited temperature variation in all directions, resulting in non-specific ice crystal morphology, and therefore in non-aligned pores (Figure 1e, top). When pectin solutions are immersed in LN, the temperature at the core of the sample remains stable before dropping sharply (243°C.min$^{-1}$). This indicates the presence of an important temperature gradient from the core of the sample to the sides. It is likely that ice crystals present in the outermost surface of the sample nucleate almost immediately when the sample is plunged into LN. Due to the radial temperature gradient, the ice crystals then grow in a radial fashion towards the center of the sample (Figure 1e, middle), which dictates the radial orientation of the final porosity. However, this gradient which strongly depends on the mold geometry cannot be easily controlled.

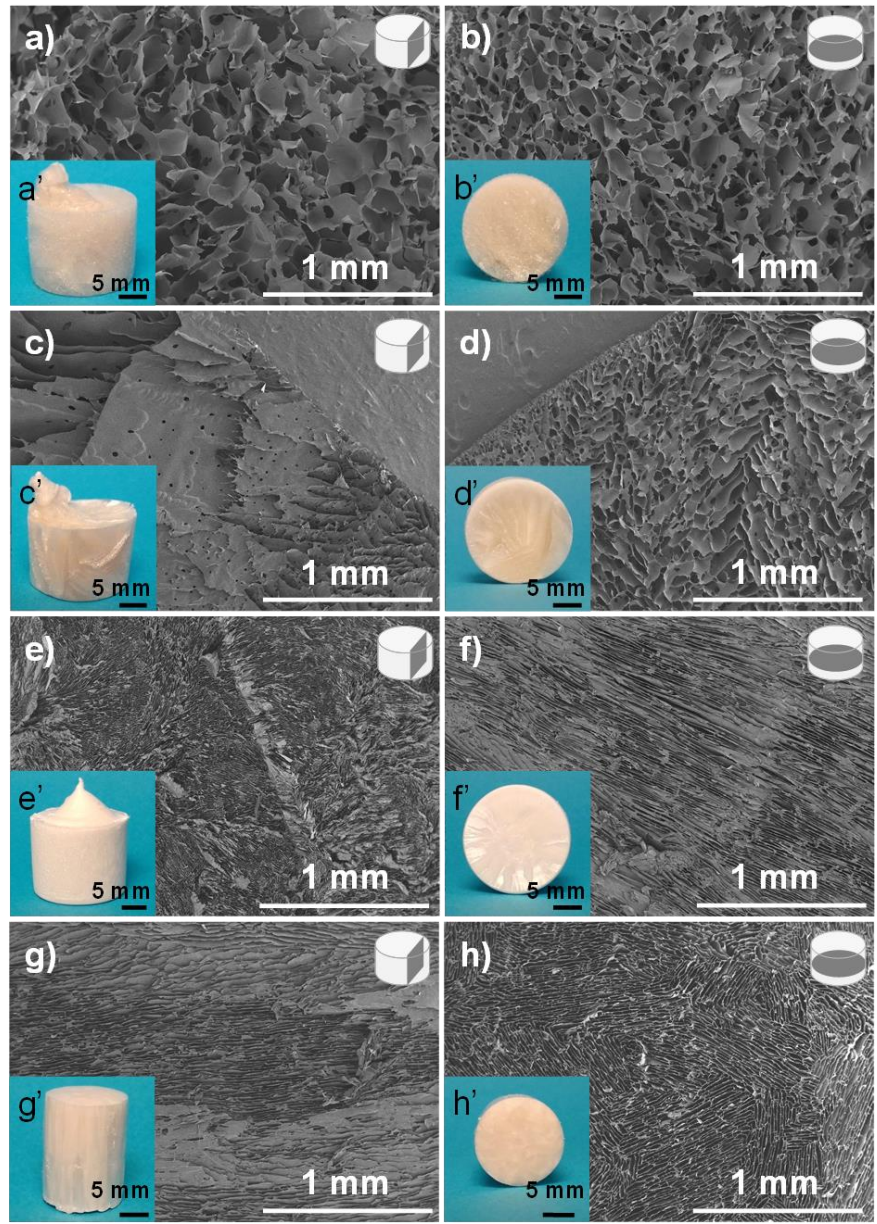

**Figure 2.** Morphology of pectin foams obtained under different freezing conditions applied on 40 gL$^{-1}$ pectin solution. SEM images of foams observed along two different planes (a-h). Insets display the macroscopic aspect of the foams obtained under the same conditions and observed along the same orientation (a'-h'). Sample preparation was conducted at -20°C freezer (a,b), -80°C freezer (c,d), LN, (e,f) and 10°C.min$^{-1}$ freeze-casting (g,h).

In contrast, the freeze-casting configuration – where the sample is placed in contact with a cooling element at the basis of the cylindrical mold – produces a longitudinal temperature gradient along the cylinder axis. The thermal gradient can be observed from the temperature difference at a given moment between the bottom and the center of the sample (Figure 1d). Thus as soon as the temperature of the cooling element reaches the transition temperature, ice crystals nucleate in the bottom region of the sample and the crystals progressively grow along the temperature gradient (Figure 1e, bottom). SEM observation of cross sections of the resulting foams reveals the presence of lamellar pores (Figure 2g), well-aligned in longitudinal channels (Figure 2h). The alignment of the pores was confirmed by the analysis of the pore orientation distribution, which also suggested the existence of domains of pores with similar orientation (Figure S2).

3.2. Liquid transport properties

The pore orientation in these freeze-cast foams is reminiscent of the aligned channels observed in wood or plant stem structures. These highly anisotropic pores are responsible for the sap transport from roots to leaves. Here, we hypothesized that the oriented porosity generated by freeze casting pectin sols can be applicable for liquid transport devices. To highlight the influence of the porosity orientation on the transport properties within pectin foams, the capillary ascension of a solution of Disperse Red in ethanol was recorded in macroporous pectin monoliths obtained in a LN bath (radial structure) and by freeze-casting (axial porosity).

A circular capillary tube is the simplest model of a porous medium. In porous media, the driving forces that results from the Laplace's depression, *i.e.* capillary forces, are relatively large because of the small size of the pores, which are generally in competition with viscous and gravitational forces. In a more classical approach [36], the height of a liquid in a vertical capillary tube of radius is obtained by solving the differential equation obtained by writing the equilibrium between these forces at any time *t*. If the porous medium is idealized as a single equivalent vertical capillary completely wetted by a given liquid, the height of advance of the meniscus *h(t)* is then given by the modified Washburn equation [37] (Eq. 1).

$$h(t) = \frac{1}{4\eta}\left[-r\xi + \sqrt{r^2\xi^2 + 8\eta(r\xi h(0) + 2\eta h^2(0) + r\sigma)t}\right] \quad \text{(Eq. 1)}$$

Where $\eta$ ($\eta = 1.17 \cdot 10^{-3}$ Pa.s) and $\sigma$ ($\sigma = 22.9$ mN/m) are, respectively, the dynamic viscosity and surface tension of the liquid [37,38]. The coefficient of friction $\xi$ is defined as the three phase line friction coefficient that describes the interactions between the solid walls and the liquid molecules [37]. By this means, we take into account the chemical nature of the substrate via the interactions at the three phase line, in terms of friction coefficient. The observed contact angle at the pectin-ethanol interface is equal to zero and the initial condition h(t=0) = h(0) = 0.

The experimental data, for the rise of ethanol in the freeze cast sample as a function of time, together with the best fit to the modified Washburn equation (Eq. 1) are presented in Figure 3 (upper graph). A good agreement between theory and experimental data is found, with $\xi = 0.045$ Pa.s and the radius of the cylindrical equivalent capillary $r = 1.935$ µm. The capillary rise of the same solution for the sample obtained by immersion into LN deviates substantially from the best fit of the previous model. The *r* parameter can be considered also as characteristic size of the porous media. The $\xi$ parameter is the expression of the dynamical character of the contact angle between the liquid and the wall of the pore. The lower graph in Figure 3 represents both the experimental and theoretical results for the foam obtained under LN freezing. The difference between the experimental and theoretical curves is most likely due to the change in dissipation regime. In the selected model, the main dissipation channel is due to the viscosity and the friction at the three-phase line (*i.e* neglecting the gravitational dissipation). The orientation of the pores is the main source of the slowing down in this experiment. As a result, it seems that the axial porosity should be favored for the conception of devices requiring high liquid mass transport capacity.

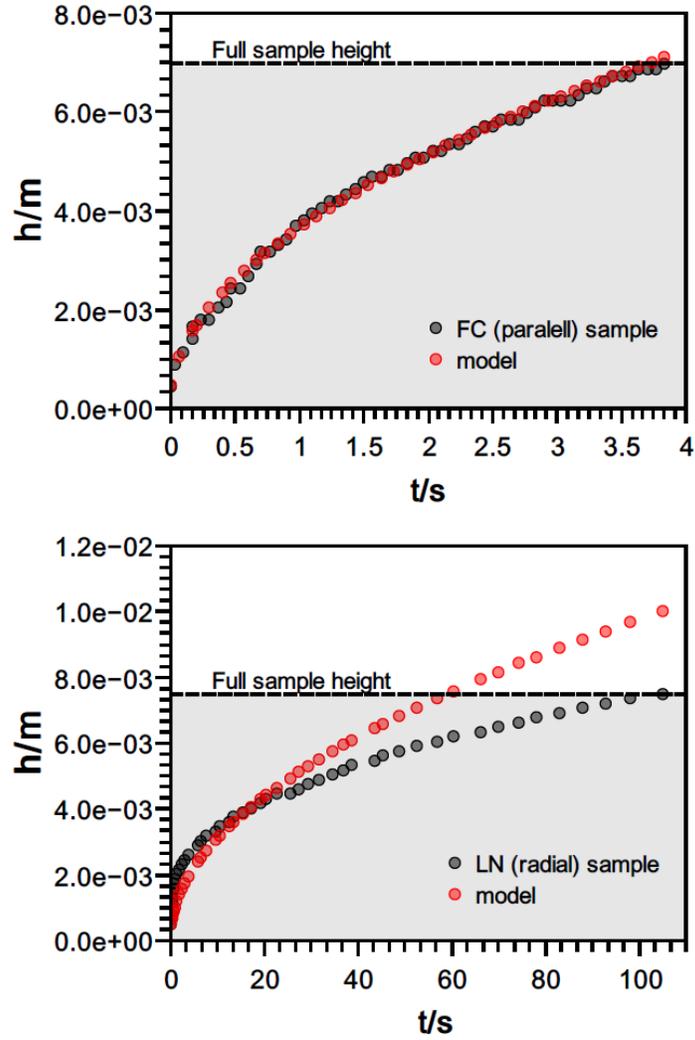

**Figure 3.** Wetting behavior of freeze-cast samples (upper graph) and liquid nitrogen (LN) frozen foams (lower graph) by Disperse Red solution in ethanol. The grey circles are extracted from the analysis of capillary rise from video footage and the red circles correspond to the best fit to the modified Washburn model. (Fit values for the bottom graph, radial LN sample are: apparent radius r =0.14 µm and $\xi$=0.05 Pa.s)

3.3. Mechanical properties

Beside the interest in terms liquid transport of such channel-like porosity, the oriented morphology obtained by freeze casting also confers an advantage from a mechanical point of view. The mechanical properties of the different materials presented earlier were assessed under compression. Two orthogonal compression directions were used on 1x1x1 cm$^3$ samples. When applicable, these directions were specifically chosen along or orthogonally to the pores.

**Table 1.** Mechanical properties and their standard deviation (sd) for the different foam morphologies and testing orientation.

|  | **Freezing Setup** | **Freezing rate / °C.min$^{-1}$** | **E / MPa** | **sd / MPa** | **σ$_Y$ / kPa** | **sd / kPa** | **Anistropy ratio** |
|---|---|---|---|---|---|---|---|
| **Compression perpendicular to the pores** | -20°C (freezer) | 1.2 | 1.1 | 0.5 | 65 | 17 | 40.7 |
|  | -80°C (freezer) | 3.6 | 0.3 | 0.1 | NA | NA | 19.4 |
|  | -196°C (LN) | 243.3 | 0.49 | 0.08 | NA | NA | 21.0 |
|  | FC at 10°C.min$^{-1}$ | 7.2 | 0.13 | 0.03 | NA | NA | 95.2 |
| **Compression along the pores** | -20°C (freezer) | 1.2 | 1.5 | 0.5 | 67 | 34 | 40.7 |
|  | -80°C (freezer) | 3.6 | 0.34 | 0.15 | NA | NA | 19,4 |
|  | -196°C (LN) | 243.3 | 0.62 | 0.78 | NA | NA | 21.0 |
|  | FC at 10°C.min$^{-1}$ | 7.2 | 2.61 | 0.65 | 92 | 2 | 95.2 |

Table 1 shows the stress/strain behavior for materials with various pore organizations (Stress/strain curves can be found in Figure S3). Compared to the other materials, the sample prepared by freeze-casting displayed a highly anisotropic behavior. When compressed in the direction of the channels, this material has a Young's Modulus of 2.8 MPa, which is typical for polymer foams of comparable apparent density [39]. However under compression orthogonal to the pores' orientation, the Young's Modulus drops to 0.13 MPa. This strong mechanical anisotropy is the direct consequence of the structural anisotropy described earlier. Macroporous materials with no specific orientation (samples obtained in freezers) have no or little mechanical anisotropy. The Young's Moduli are in the same order of magnitude in both directions (between 1.1 and 1.5 MPa for the -20°C freezing *vs* 0.3 MPa for the -80°C freezing). It is however important to note that the value of the Young's modulus differs greatly between these two samples. This might be attributed to the slightly elongated shape and more lamellar structure of the material prepared at -80°C. The radial structure (samples obtained in liquid nitrogen), despite its specific pore orientation, displays little or no mechanical anisotropy. This may be due to the fact that the compression tests are not performed directly along the pores direction but rather along the diameter of the cylindrical mold. As a consequence, the measured Young's Modulus corresponds in fact to an average effect dictated by the distribution of the pore directions. These results can be interpreted according to simple composite models describing the elastic behavior of matrices (or continuous phase) reinforced with an anisotropic reinforcing phase. In porous materials the specificity is that the reinforcing phase can be modeled as an inclusion with an elastic modulus equal to zero. According to the relative orientation between the materials' pores and the compression axis the resulting material elastic properties can be describes according to two extreme models: the rule of mixtures (ROM) and the inverse rule of mixtures (IROM). If the inclusions (pores) are aligned along the compression axis, then the mechanical behavior can be roughly described by the ROM and its modulus will be maximal. On the contrary, if the pores are oriented perpendicular to the compression direction then the material compressive modulus will be roughly described by the IROM, where the elastic modulus depends strongly on the lowest modulus among the matrix/inclusion couple.

These results prove that a precise control over the freezing conditions (*i.e.* the cooling profiles) confers control over the morphology of materials obtained by ice-templating, which has a direct influence on macroscopic properties such as the mechanical behavior. As in most of higher plants, where the xylem is oriented along the main mechanical solicitation direction, the morphology of freeze cast pectin samples displays a co-alignment between the most efficient liquid transport channels and the orientation exhibiting the highest mechanical resistance to compression.

3.4 Influence of pectin concentration

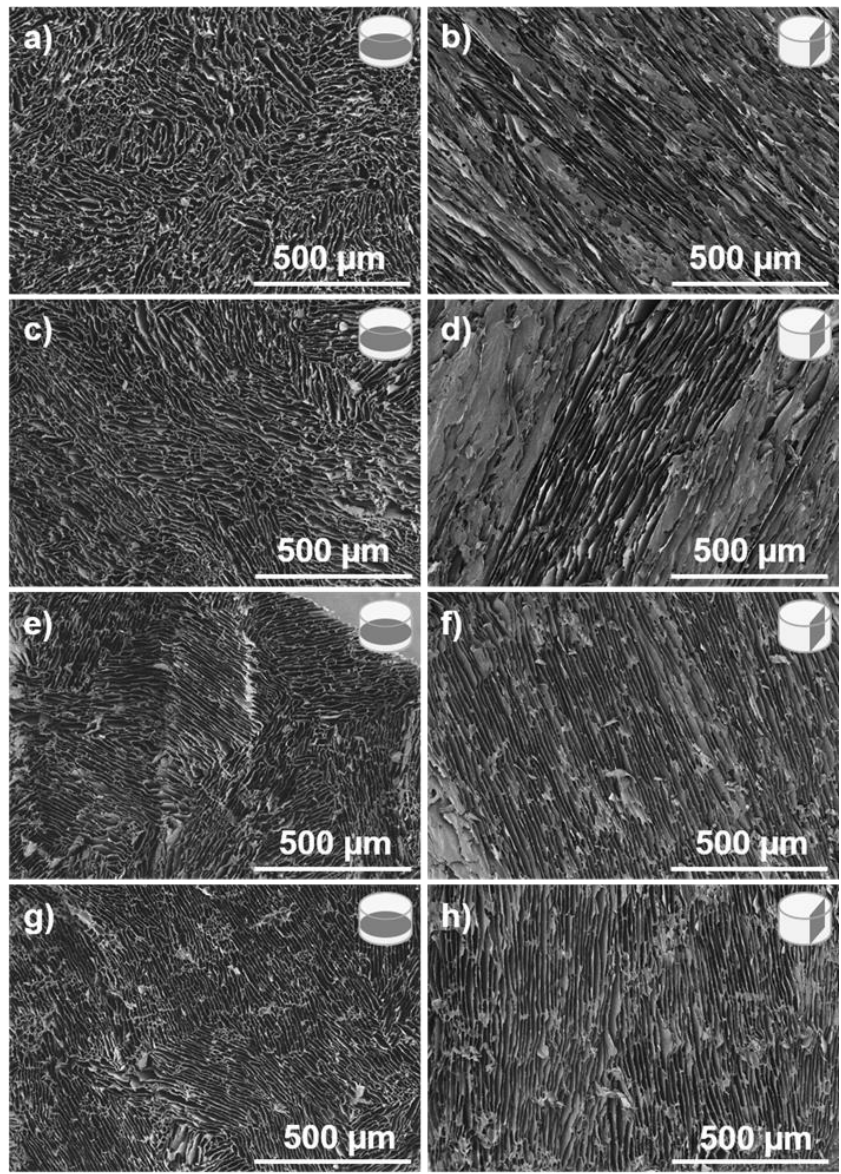

**Figure 4.** SEM of sugar beet pectin prepared at different initial concentrations, scale bar is 500 µm. a) and b) 20 g.L$^{-1}$; c) and d) 30 g.L$^{-1}$; e) and f) 40 g.L$^{-1}$; g) and h) 50 g.L$^{-1}$. Left hand-side images are taken along the ice growth direction, right-hand side images were taken perpendicular to the ice growth direction.

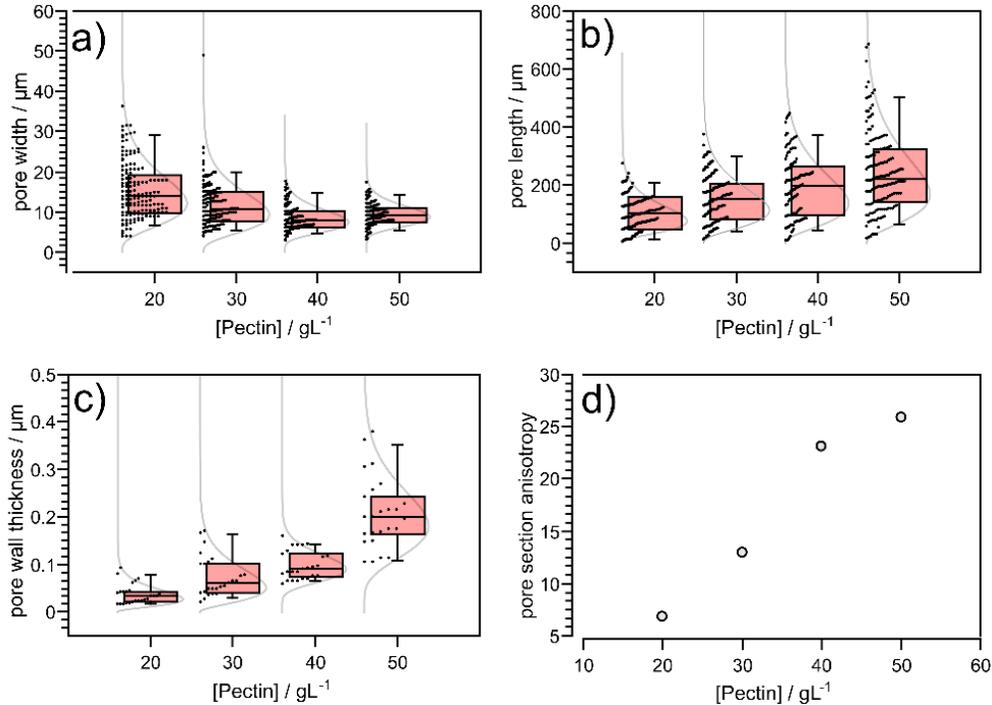

**Figure 5.** Influence of pectin initial concentration on the pore morphology in freeze cast foams. The boxplots a) to c) represent the different dimension of the macroporous features. The individual data points (black dots) used for each measurement along with their best fit to a gamma function (light grey line) are displayed next to the corresponding boxplots. Moustaches delimit the 5 and 95 percentile of the distribution and box limits represent the 1st and 3rd quartiles. The dependence between the average pore section anisotropy and the pectin concentration is plotted in d).

While freezing conditions allows for a precise control over the morphology, liquid transport behavior and mechanical properties of pectin foams, other parameters, such as the initial pectin concentrations may also impact the final porous structure. Foams prepared from aqueous pectin solutions ranging from 20 to 50 g.L$^{-1}$ all display similar, well-aligned pores (Figure 4 b, d, f and h). The appearance of longitudinal pores is mostly controlled by the thermal gradient developed throughout the sample during the freezing step, which is in all likelihood is barely affected by the pectin concentration. However, cross section SEM images (Figure 4a, c, e and g) display notable differences depending on the pectin concentration.

The impact of pectin solution concentration on ice growth that can be evaluated from the size, morphology and organization of the pores (Figure 5 and Figure S4). Pectin foams obtained by freeze casting a solution prepared at 20 g.L$^{-1}$ display a relatively large distribution of pores with little anisotropy in its cross-section. The average pore width (wall-to-wall distance) is centered at 15 µm and the cross-section length of the pores averages at ca. 100 µm. When pectin foams are prepared from solutions with increasing concentrations (up to 50 g.L$^{-1}$), the pore width size decreases to 8.5 µm and their distribution becomes narrower (between 20 gL$^{-1}$ and 50 gL$^{-1}$ the standard deviation decreases from 7 µm to 2.6 µm, respectively). Inversely, the average pore length increases up to 245 µm. These morphological features correlate with the formation of thicker pore walls, from 39 nm to 208 nm at 20 g.L$^{-1}$ and 50 g.L$^{-1}$, respectively. Overall, the increase in pectin concentration favors the formation of pores with smaller size distribution, higher anisotropy (Figure 5d) and more effective packing of pores in ordered domains (Figure S4). These consequences are coherent with the higher viscosity of the pectin solutions and how it may affect the growth of ice crystals.

Tuning the initial biopolymer concentration translates into variations of the final density of the materials and therefore provides a simple way to tailor their mechanical properties (Figure 6 and Table S2). Young's Modulus measured for sugar beet pectin freeze cast foams are between 1 and 5 MPa, which are typical for polymer foams with density ranging from 20 kg.m$^{-3}$ to 50 kg.m$^{-3}$ [40]. The stress-strain profiles show typical behavior for polymer foams, with an initial elastic behavior (up to about 7% strain), followed by a plastic deformation plateau (between 7 and 40% strain). The material then undergoes a densification, which translates in a strain increase beyond 40% deformation. Foams with different densities follow the same general profile, however the yield strength changes significantly (variation between 20 and 186 kPa), which can be linked to the changes in pore walls thickness previously mentioned.

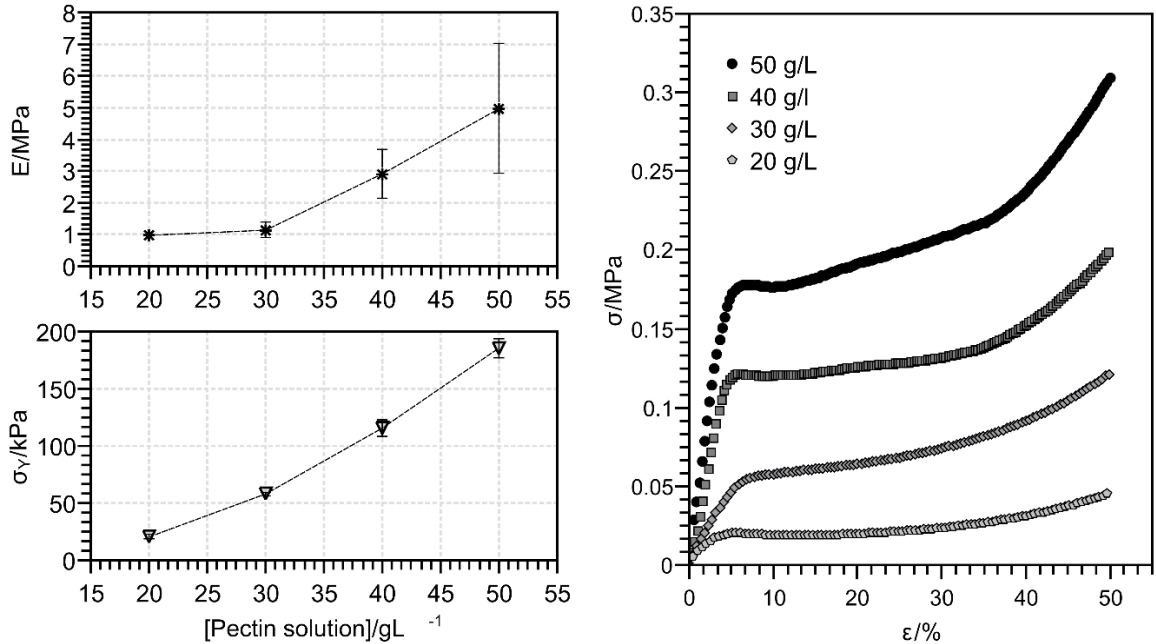

**Figure 6.** Influence of pectin initial concentration on the mechanical properties of freeze cast foams. The Young's modulus (upper left panel) and yield stress (lower left panel) were inferred from the stress-train diagram obtained under compression of 1x1x1 cm$^3$ foams along the pores' longitudinal axis.

Compared to previously reported pectin-based materials [41,42], freeze cast sugar beet pectin is particularly efficient in terms of the specific moduli, $E_s$,(*i.e.* the elastic modulus divided by the materials' apparent density) at bulk density below 50 kg.m$^{-3}$, ranging up to 107 kPa.m$^3$.kg$^{-1}$. In Figure 7, sugar beet pectin compressive modulus measured along the pores was plotted side by side with the literature values for other pectin-based foams (Figure 7). From the plot, freeze cast sugar beet pectin foams' modulus are similar to citrus-based aeropectin but with significantly lower bulk densities. When compared with freeze-cast citrus pectin, sugar beet foams can display compressive moduli over one order of magnitude larger at similar bulk densities. Such an increased efficiency may arise from the higher aspect ratio and more controlled orientation of the porosity of the freeze cast sugar beet foams as compared with the radial orientation of the pores described by Rudaz [42] or the low anisotropy pores generated by the gelation/coagulation method [41].

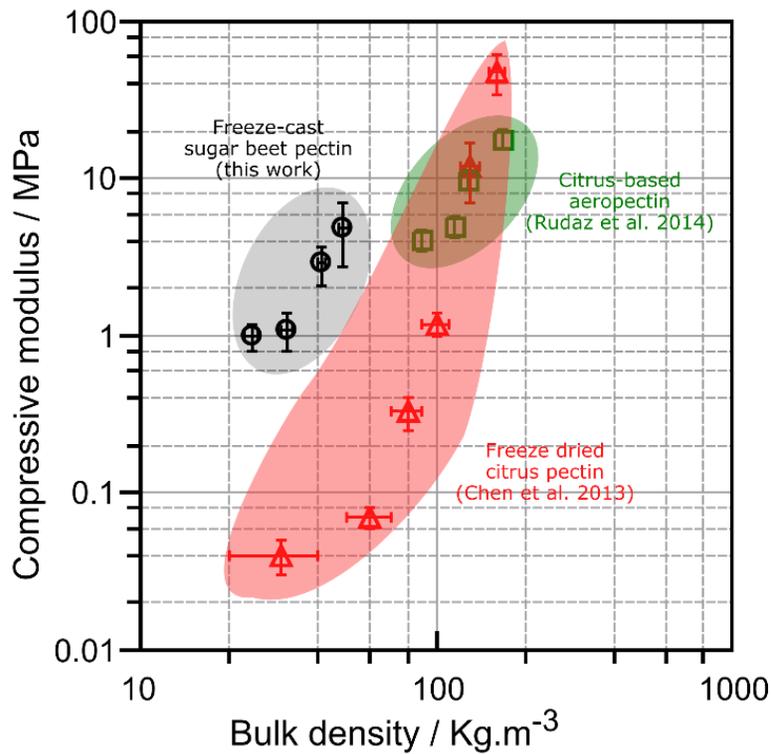

**Figure 7.** Ashby plot of sugar beet pectin prepared by freeze casting in this work (black circles) and literature values for other lightweight materials prepared from citrus pectin (red triangles: freeze dried [41]; green squares: supercritical-dried aerogels [42]).

**Conclusions**

Freeze casting of sugar beet pectin solutions provides an easy and efficient way to prepare lightweight materials with high specific moduli and efficient liquid transport properties. The foams' properties are highly dependent on the processing parameters that allow for a precise control of the porous materials morphology. Other parameters, including mold geometry or presence of additive could be investigated to further widen the range of possible porous characteristics. Multi-step processing methods may also be devised to optimize the macroscopic properties of these materials. In a more general context, the design of all-polysaccharide lightweight materials has so far been mainly focused on the development of composite materials combining reinforcing moieties such as cellulose and its derivatives dispersed in other polysaccharide matrices. Here we show that structuring of a single polysaccharide material by freeze casting is possible and enables the upcycling of sugar beet pectin, a particularly undervalued polysaccharide. Extension to other water-soluble polysaccharides may prove instrumental in designing new materials out of undervalued resources.

**Acknowlegments**

This work was supported by French state funds managed by the ANR within the *Investissements d'Avenir* program under reference ANR-11-IDEX-0004-02, and more specifically within the framework of the Cluster of Excellence MATISSE led by Sorbonne Universités. M. Turmine is warmly acknowledged for the measurement of the liquid surface tension.